\documentclass[twocolumn,aps,prl,preprintnumbers,superscriptaddress]{revtex4}
\usepackage{}
\usepackage{bbm}
\usepackage[latin9]{inputenc}
\usepackage{amsmath}
\usepackage{amssymb}
\usepackage{graphicx}
\usepackage{mathrsfs}
\usepackage{amsfonts}
\usepackage{amsthm}
\usepackage{color}
\usepackage{txfonts}
\usepackage{lipsum}
\usepackage{gensymb}
\usepackage[colorlinks=true,citecolor=blue,linkcolor=blue,urlcolor=blue,anchorcolor=blue]{hyperref}%
\hypersetup{colorlinks=true,citecolor=blue,linkcolor=blue,urlcolor=blue}
\providecommand{\U}[1]{\protect\rule{.1in}{.1in}}
\makeatletter

\newcommand{\Rmnum}[1]{\expandafter\@slowromancap\romannumeral #1@}
\makeatother
\makeatletter
\@ifundefined{textcolor}{}
{
\definecolor{BLACK}{gray}{0}
\definecolor{WHITE}{gray}{1}
\definecolor{RED}{rgb}{1,0,0}
\definecolor{GREEN}{rgb}{0,1,0}
\definecolor{BLUE}{rgb}{0,0,1}
\definecolor{CYAN}{cmyk}{1,0,0,0}
\definecolor{MAGENTA}{cmyk}{0,1,0,0}
\definecolor{YELLOW}{cmyk}{0,0,1,0}
}
\makeatother
\begin{document}
\title{Achieving unidirectional propagation of twisted magnons in a magnetic nanodisk array}
\author{Zhixiong Li}
\affiliation{School of Physics, Central South University, Changsha, 410083, China}
\author{Xiansi Wang}
\affiliation{School of Physics and Electronics, Hunan University, Changsha, 410082, China}
\author{Xuejuan Liu}
\affiliation{College of Physics and Engineering, Chengdu Normal University, Chengdu 611130, China}
\affiliation{School of Physics and State Key Laboratory of Electronic Thin Films and Integrated Devices, University of Electronic Science and Technology of China, Chengdu 610054, China}
\author{Peng Yan}
\email[Corresponding author: ]{yan@uestc.edu.cn}
\affiliation{School of Physics and State Key Laboratory of Electronic Thin Films and Integrated Devices, University of Electronic Science and Technology of China, Chengdu 610054, China}

\begin{abstract}
Twisted magnons (TMs) have great potential applications in communication and computing owing to the orbital angular momentum (OAM) degree of freedom. Realizing the unidirectional propagation of TMs is the key to design functional magnonics devices. Here we theoretically study the propagation of TMs in one-dimensional magnetic nanodisk arrays. By performing micromagnetic simulations, we find that the one-dimensional nanodisk array exhibits a few bands due to the collective excitations of TMs. A simple model by considering the exchange interaction is proposed to explain the emerging multiband structure and theoretical results agree well with micromagnetic simulations. Interestingly, for a zigzag structure, the dispersion curves and propagation images of TMs show obvious nonreciprocity for specific azimuthal quantum number ($l$), which originates from a geometric effect depending on the phase difference of TMs and the relative angle between two adjacent nanodisks. Utilizing this feature, one can conveniently realize the unidirectional propagation of TMs with arbitrary nonzero $l$. Our work provides important theoretical references for controlling the propagation of TMs.

\end{abstract}
\maketitle
\section{Introduction}
Ever since the quantized orbital angular momentum (OAM) states were originally introduced in photonics \cite{AllenPRA1992,MolinaNP2007,PadgettOE2017,JiS2020,FrankeNRP2022}, the peculiar twisted structure has been rapidly extended to a broad field of electronics \cite{UchidaN2010,VerbeeckN2010,McmorranS2011,SilenkoPRL2017,LloydRMP2017}, acoustics \cite{DashtiPRL2006,AnhauserPRL2017,BareschPRL2018,MarzoPRL2018,BliokhPRB2019}, neutronics \cite{ClarkN2015,CappellettiPRL2018,LarocqueNP2018,AfanasevPRC2019,SherwinPLA2022}, and spintronics \cite{JiangPRL2020,JiaNC2019,ChenAPL2020,JiaJO2019,LiPRB2022}. In magnetic system, the magnons (quantized quasiparticle of spin wave) carrying OAM are called twisted magnons (TMs) \cite{JiangPRL2020,JiaNC2019}. The researches about the OAM states of magnons have attracted growing interest owing to both the fundamental interest and potential applications. By using the twisted phase structure of TMs as individual information channels, it is possible to realize the frequency-division multiplexing which can greatly enhance the communication capacity of magnons \cite{LiPRB2022}. It has been proposed that the TMs can act as ``magnetic tweezers" to drive the rotation of spin texture (such as skyrmion) \cite{JiangPRL2020}. Very recently, Wang \emph{et al.} \cite{WangPRL2022} showed that the magnonic frequency comb emerges in the nonlinear interaction between TMs and magnetic vortex.

Twisted magnonics focuses on the generation, propagation, manipulation, and detection of TMs. To design various functional devices based on TMs, realizing the unidirectional propagation of TMs with arbitrary OAM quantum number ($l$) is one of the central tasks. Generally speaking, TMs only exist in magnetic nanocylinder and nanodisk. On the one hand, Jiang \emph{et al.} \cite{JiangPRL2020} and Jia \emph{et al.} \cite{JiaNC2019} have theoretically studied the spectrum of TMs in a single magnetic nanocylinder. In such configuration, it is however difficult to excite the TMs with a specific $l$ because of the multiband structure. On the other hand, although the intrinsic dynamics of TMs in single magnetic nanodisk has been investigated \cite{LiPRB2022}, the collective dynamics of TMs in nanodisk arrays is rarely explored. The magnetic nanodisk array is an ideal platform for studying the collective propagation of TMs with the following reasons: (i) 
\begin{figure}[!htbp]
\begin{centering}
\includegraphics[width=0.42\textwidth]{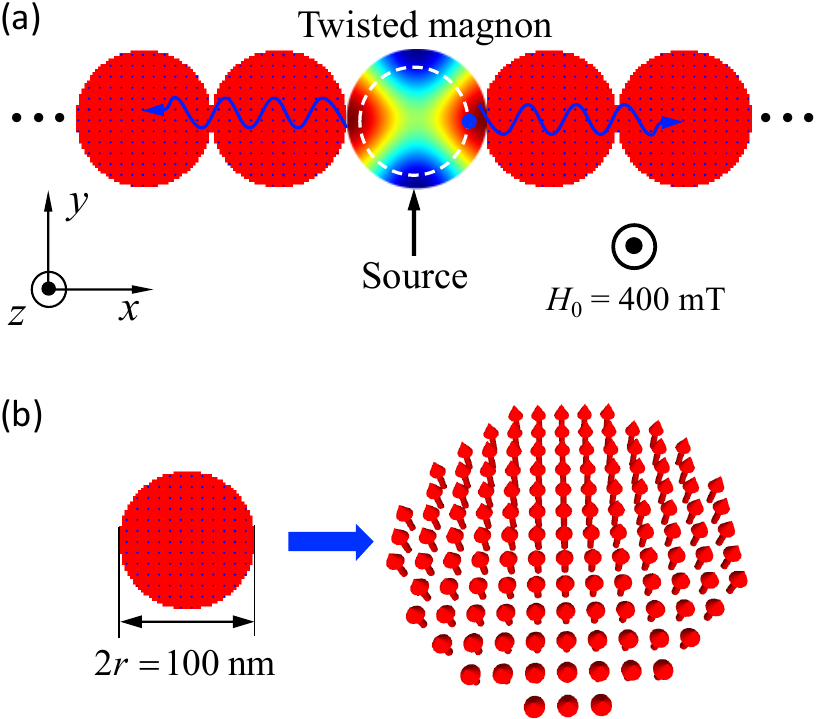}
\par\end{centering}
\caption{(a) The illustration of a straight one-dimensional lattice containing 101 magnetic nanodisks. A uniform static magnetic field is applied along the $z-$axis to perpendicularly magnetize the magnetic moments. The TMs are excited at the center disk, marked by black arrow. (b) Zoomed in details of one nanodisk. The red arrows denote the local magnetization.}
\label{Figure1}
\end{figure}The desired lattice structure based on magnetic nanodisks can be fabricated within the reach of current experimental techniques, for example, electron-beam lithography \cite{HanSR2013,BehnckePRB2015,SunPRL2013}. (ii) It is convenient to excite TMs with arbitrary $l$ in nanodisk arrays by means of the so-called spin-to-orbital angular momentum conversion mechanism \cite{LiPRB2022}. (iii) For two- or three-dimensional nanodisk lattice, one may realize the chiral propagation of TMs with topological features. It is thus naturally expected that the collective excitations of TMs in nanodisks array can exhibit abundant physics (unidirectional propagation for instance), which should provide important theoretical references for designing functional magnonic devices.      

In this work, we study the collective dynamics of TMs in one-dimensional magnetic nanodisk arrays. For a straight lattice, the system supports a few symmetric magnon bands describing different collective excitation modes of TMs. A simple exchange model is proposed to explain the emergence of multiband structure. Interestingly, for the zigzag structure, the TM dispersion relations can exhibit visible nonreciprocity. These asymmetric bands are explained by a geometric effect: when the phase difference of TMs does not match the geometric angle ($\theta$) [see Fig. \ref{Figure3}(b)], the nonreciprocity occurs. It allows us to realize unidirectional propagation of TMs for any nonzero $l$ by tuning $\theta$. In addition, we find that the propagation direction of TMs can be conveniently tuned by changing the sign of $l$ or the position of excitation field. Our results provide a simple and effective method to control the propagation of TMs which should greatly promote the development of twisted magnonics.

The paper is organized as follows. In Sec. \ref{section2}, we present micromagnetic simulations for collective excitations of TMs in straight one-dimensional nanodisk lattices. Section \ref{section3} introduces the theoretical model to explain the emerging multiband structures of TMs. In Sec. \ref{section4}, we focus on the unidirectional propagation of TMs in a zigzag nanodisk array. Discussion and conclusion are drawn in Sec. \ref{section5}.

\section{Micromagnetic simulation}\label{section2}
We consider a straight one-dimensional lattice consisting of 101 identical magnetic nanodisks with radius $r=50$ nm and thickness $d=2$ nm, as shown in Fig. \ref{Figure1}. The distance between nearest-neighboring nanodisks is $2r$, which indicates that the TMs can interact with each other through the exchange interaction. The material parameters of yttrium iron garnet (YIG) are used \cite{LiPRB2022}: the saturation magnetization $M_{s}=1.92\times10^{5}$ Am$^{-1}$, the exchange stiffness $A=3.1\times10^{-12}$ Jm$^{-1}$, and the Gilbert damping constant $\alpha=10^{-3}$. The magnetic moments are perpendicularly magnetized by external magnetic field $H_{0}=400$ mT. The cell size is set to be $2\times2\times2$ nm$^{3}$. The micromagnetic software package MUMAX3 \cite{VansteenkisteAA2014} is used to simulate the magnetization dynamics. To excite the collective oscillation of TMs, we apply a sinc-function magnetic field  
\begin{equation}\label{Eq1}
\mathbf{H}(t)=H_{1}\frac{\text{sin}[2\pi f_{0}(t-t_{0})]}{2\pi f_{0}(t-t_{0})}[\text{cos}(l\phi),\text{sin}(l\phi),0],
\end{equation}  
with $H_{1}=40$ mT, $f_{0}=15$ GHz (cutoff frequency), and $t_{0}=1$ ns, confined to the disk located at the center of the lattice, as labeled by the black arrow in Fig. \ref{Figure1}(a). Here $\phi$ is the polar angle. The spatiotemporal profile of magnetizations in all nanodisks are recorded every 20 ps and the total simulation time is 200 ns.

The dispersion relation of TMs is obtained by calculating the spatiotemporal fast Fourier transformation (FFT) of the averaged (over the whole disk) magnetization $x-$component $\langle m_{x}\rangle$ (or $y-$component). For every azimuthal (OAM) quantum
\begin{figure}[!htbp]
\begin{centering}
\includegraphics[width=0.44\textwidth]{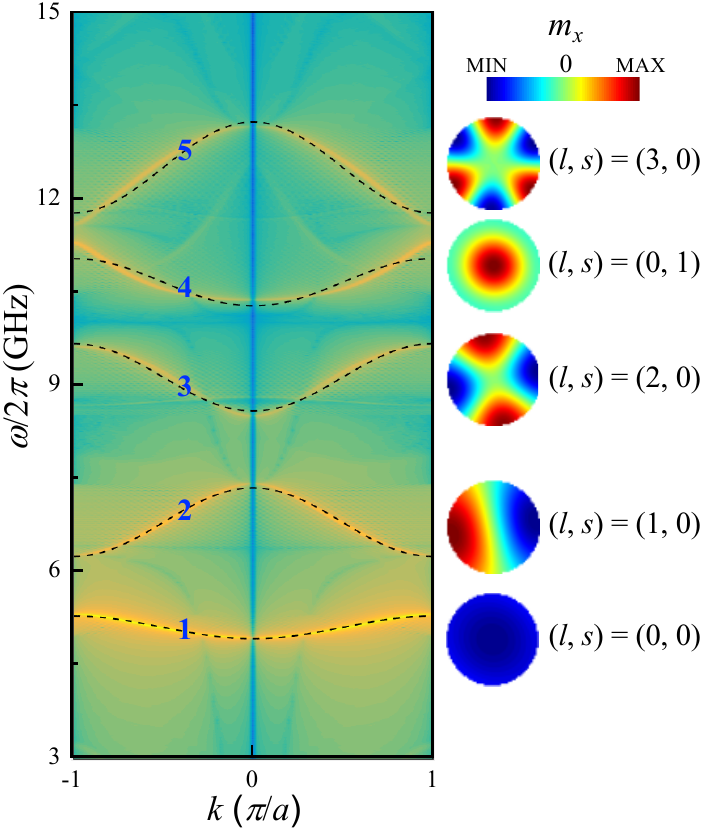}
\par\end{centering}
\caption{Left column: dispersion relations of TM for the structure in Fig. \ref{Figure1}(a). The background color and dashed black lines represent the simulation results and analytical formulas, respectively. Right column: the spatial distribution of FFT intensity for different TM modes which correspond to the five bands emerging in left column. Here $s$ denotes the radial quantum number which counts the number of nodes along the radial direction.}
\label{Figure2}
\end{figure}number $l$, we can calculate the spectrum. To get the full band structure, we sum the spectra for all $l$. Figure \ref{Figure2} shows the results, from which we can clearly see that the system exhibits five separate dispersion curves below 15 GHz, as marked by blue arabic number 1-5. Besides, by analyzing the spatial distribution of the FFT intensity for these bands, we can identify five different TMs modes, as shown in right column of Fig. \ref{Figure2}. For each dispersion relations, at the bottom (top) of bands, the adjacent TMs oscillate in-phase (out-of-phase), which is similar to other (quasi-)particles system. Interestingly, we find that the signs of the group velocity are opposite when $l$ is even (bands 1, 3, and 4) and odd (bands 2 and 5) for the same value of wave vector $k$.

\section{Theoretical model}\label{section3}
To explain the emerging multiband structure of TM, we propose a theoretical model which is similar to the framework of massless Thiele's equation \cite{ThielePRL1973}. Here the dynamics of TM can be described by analogous Thiele's equation based on the following facts. At first, due to the distinctive mode profile of TM (see Fig. \ref{Figure2}), it is reasonable to use a wavepacket description. Then we consider the position of the peak (or trough) to represent the TM in nanodisk because of the circular symmetry, as denoted by blue ball in Fig. \ref{Figure1}(a) [here $(l,s)=(2,0)$]. At last, we envision that the steady-state magnetization of the nanodisk only depends on the position of TM.

Assuming the displacement vector of TM from the disk center in $j$th nanodisk as $\mathbf{U}_{j}=(u_{j},v_{j})$, we obtain the dynamic equation characterizing TM as
\begin{equation}\label{Eq2}
G\hat{z}\times\frac{d\mathbf{U}_{j}}{dt}+\mathbf{F}_{j}=0,
\end{equation}     
where $G$ is a gyroscopic coefficient depending on both the values $l$ and $s$. The conservative force can be expressed as $\mathbf{F}_{j}=-\partial W/\partial\mathbf{U}_{j}$. Here $W$ denotes the total potential energy
\begin{equation}\label{Eq3}
W=\sum_{j}K\mathbf{U}_{j}^{2}/2+W_{d}+W_{z}+W_{e}.
\end{equation}   
The first term at the right hand in Eq. \eqref{Eq3} originates from the confinement of disk boundary, while the terms $W_{d}$, $W_{z}$, and $W_{e}$ represent the potential energy from magnetostatic, Zeeman and exchange interactions, respectively.

Then we consider the excitation of TM, i.e., $\mathbf{m}=(m_{x},m_{y},1)$ with $m_{x}^{2}+m_{y}^{2}\ll 1$, with $\mathbf{m}$ being the unit vector of the local magnetic moment. On the one hand, it is straightforward that the Zeeman energy -$\mu_{0}M_{s}\int H_{0}\hat{z}\cdot\mathbf{m}d\mathbf{r}=-N\mu_{0}M_{s}$ is a constant. Here, $\mu_{0}$ is vacuum permeability and $N$ is the number of magnetic moment. On the other hand, the magnetostatic energy can also be treated as constant value under the linear approximation (see Appendix A for details). At last, we assume that the exchange energy takes the simple form \cite{LiPRB2021}
\begin{equation}\label{Eq4}
W_{e}=\sum_{k\in\langle j \rangle}I\mathbf{U}_{j}\cdot\mathbf{U}_{k},
\end{equation}
where $I$ is the coupling coefficient. Then the total potential energy becomes the following form
\begin{equation}\label{Eq5}
W=W_{0}+\sum_{j}K\mathbf{U}_{j}^{2}/2+\sum_{k\in\langle j \rangle}I\mathbf{U}_{j}\cdot\mathbf{U}_{k},
\end{equation}  
where $W_{0}=W_{d}+W_{z}$ denotes the constant term of energy, $\langle j\rangle$ is the set of nearest neighbors of $j$.


Substituting Eq. \eqref{Eq5} into Eq. \eqref{Eq2} and assuming $\psi_{j}=u_{j}+iv_{j}$, we obtain the eigen-equation
\begin{equation}\label{Eq6}
\frac{d\psi_{j}}{dt}+iC_{1}\psi_{j}+iC_{2}(\psi_{j-1}+\psi_{j+1})=0,
\end{equation}  
with parameters $C_{1}=K/G$ and $C_{2}=I/G$. Then we consider the plane-wave expansion of $\psi_{j}=\phi_{j}\text{exp}(-i\omega t)\text{exp}[i(n\mathbf{k}\cdot\mathbf{a})]$, where $\mathbf{k}$ is the wave vector, $n$ is an integer, and $\mathbf{a}=a\hat{x}$ is the 
\begin{table}[htbp]
\caption{The fitting parameters for different TM modes}
\label{tab1}
\centering
\begin{tabular}{cccc ccc}
\hline\hline\noalign{\smallskip}
TM mode $(l,s)$\,\,\,\, & \,\,\,\,$(0,0)$\,\,\,\, & \,\,\,\,$(1,0)$\,\,\,\, & \,\,\,\,$(2,0)$\,\,\,\, & \,\,\,\,$(0,1)$\,\,\,\, & \,\,\,\,$(3,0)$\,\,\,\,  \\
\noalign{\smallskip}\hline\noalign{\smallskip}
$C_{1}/2\pi$ (GHz) & 5.11& 6.80 & 9.13  & 10.66 & 12.51\\
$C_{2}/2\pi$ (GHz) & -0.18& 0.55 & -0.54 & -0.38 & 0.73\\
\noalign{\smallskip}\hline\noalign{\smallskip}
\end{tabular}
\end{table}basis vector with $a=100$ nm representing the lattice constant. We thus obtain the dispersion relation of TM
\begin{equation}\label{Eq7}
\omega=C_{1}+2C_{2}\text{cos}(\mathbf{k}\cdot\mathbf{a}).
\end{equation}

Then we use the formula \eqref{Eq7} to fit the dispersion curves of TM obtained from micromagnetic simulation. The dashed black lines in Fig. \ref{Figure2} shows the best fit of the numerical data, from which we can clearly see that the theoretical curves agree well with simulations for small $l$ or $s$ (bands 1, 2, and 3). However, for larger values of $l$ or $s$ (bands 4 and 5), there exists obvious discrepancy between theoretical value and micromagnetic result, which may come from the fact that the form of exchange energy \eqref{Eq4} is too simple to accurately describe the interaction between TMs with high $l$ (or $s$). The fitting parameters $C_{1}$ and $C_{2}$ for different $l$ or $s$ are summarized in Table \ref{tab1}. Overall, $C_{1}$ and $C_{2}$ are sensitive to $l$ and $s$: (i) The parameter $C_{1}$ is always positive, while $C_{2}$ is negative (positive) when $l$ takes an even (odd) number. (ii) With the increase of $l$ ($s$) for fixed $s$ ($l$), the magnitude of $C_{1}$ and $C_{2}$ increases.    

\begin{figure}[!htbp]
\begin{centering}
\includegraphics[width=0.48\textwidth]{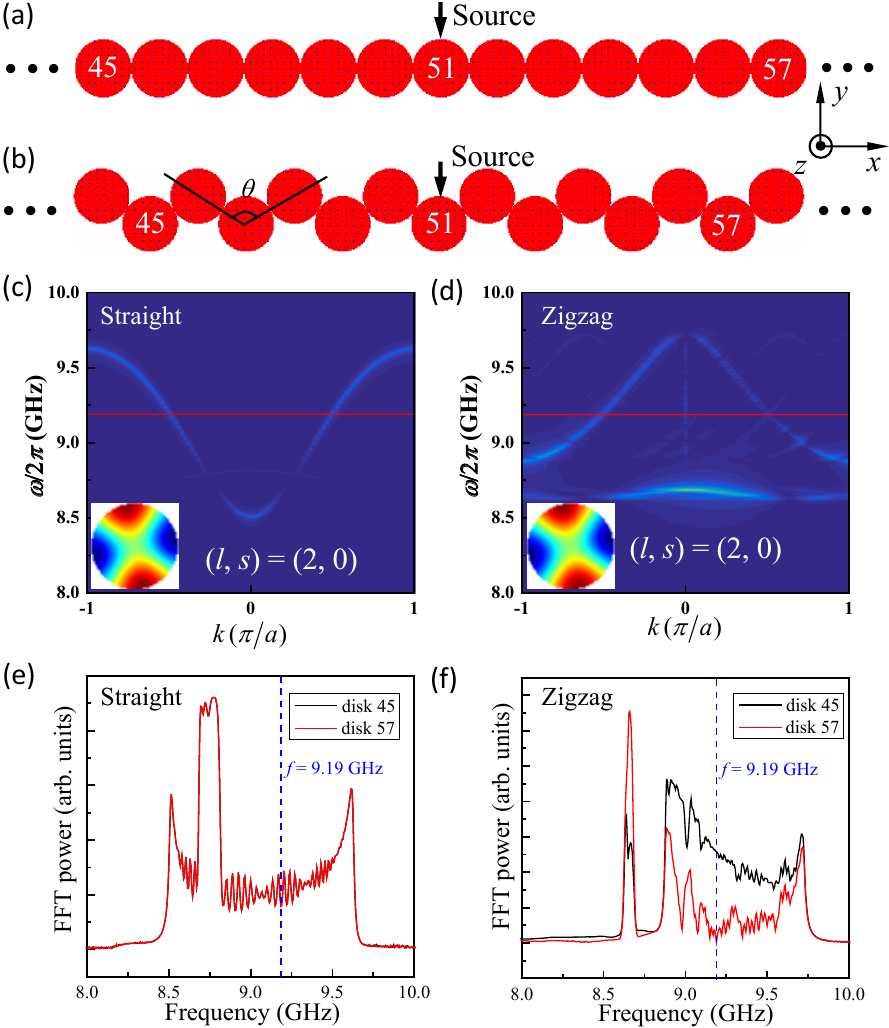}
\par\end{centering}
\caption{Schematic diagrams of (a) straight and (b) zigzag one-dimensional nanodisk array. Black arrows denote the disks where the excitation fields are applied. The angle $\theta=2\pi/3$ (formed by lines connecting the centers of two adjacent disks) describes the geometric shape of zigzag structure. The band structures for (c) straight and (d) zigzag structure with $(l,s)=(2,0)$. The red lines represent the frequency $\omega/2\pi=9.19$ GHz. The temporal Fourier spectra of the magnetization oscillation at disks 45 and 57 [as marked in Figs. \ref{Figure3}(a) and (b)] for (e) straight and (f) zigzag lattice.}
\label{Figure3}
\end{figure}

\section{Unidirectional propagation of TMs}\label{section4} 

\begin{figure}[!htbp]
\begin{centering}
\includegraphics[width=0.48\textwidth]{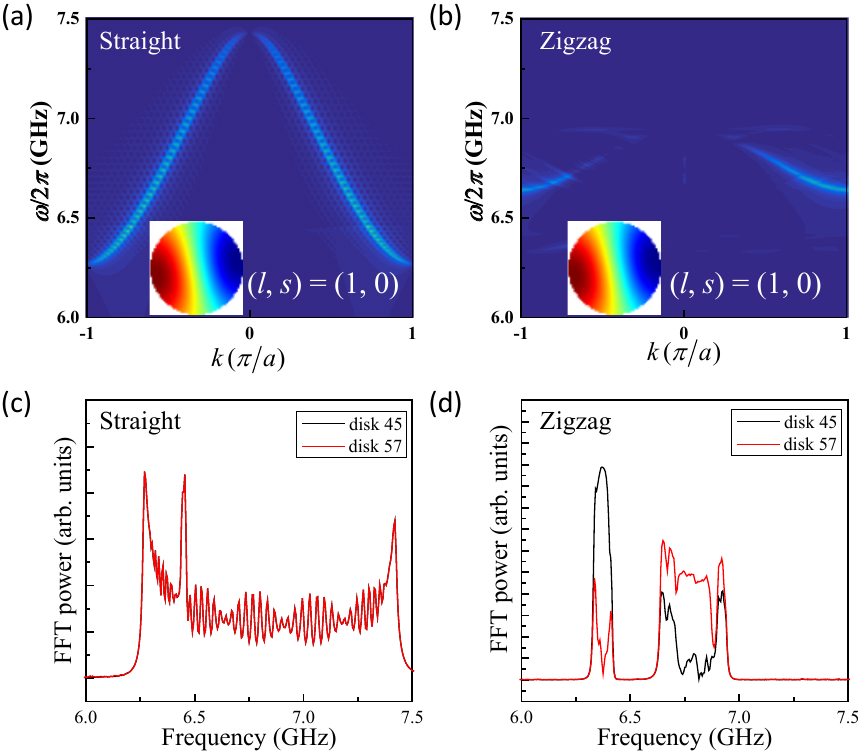}
\par\end{centering}
\caption{The band structures for (a) straight and (b) zigzag lattice with $(l,s)=(1,0)$. The corresponding spectra of the magnetization oscillation at disks 45 and 57 for (c) straight and (d) zigzag lattice.}
\label{Figure4}
\end{figure}

Next, we discuss the propagation characteristics of TMs in zigzag structure, as shown in Fig. \ref{Figure3}(b). By changing the value of $\theta$, one can tune the geometric shape of the lattice. We first choose $\theta=2\pi/3$ as an example. Interestingly, in this case, the dispersion relations of TM show obvious nonreciprocity for $l=1$ and $l=2$ [see Figs. \ref{Figure3} and \ref{Figure4}], which is in a sharp contrast to the straight structure. We focus on this feature in this section.

Figures \ref{Figure3}(c) and \ref{Figure3}(d) show the band structures of TM with $(l,s)=(2,0)$ for straight and zigzag lattice, respectively. Here the excitation fields with the form of Eq. \eqref{Eq1} are applied to disk 51 (the center disk). One can clearly see that the FFT strength of dispersion curves are symmetric for $+k$ and $-k$ in the straight lattice, while it shows visible asymmetric feature for the zigzag case. Besides, we plot the spectra of the magnetization ($m_{x}$) oscillation at disks 45 and 57 [as marked in Figs. \ref{Figure3}(a) and \ref{Figure3}(b)], as shown in Figs. \ref{Figure3}(e) and \ref{Figure3}(f), from which one can identify again the existence of nonreciprocity for TMs propagation in zigzag lattice. What's more, the TM with $(l,s)=(1,0)$ also exhibits the similar behaviors, as plotted in Fig. \ref{Figure4}. The band structures [Figs. \ref{Figure4}(a) and \ref{Figure4}(b)] and the disk spectra [Figs. \ref{Figure4}(c) and \ref{Figure4}(d)] clearly show that the propagation of TMs [$(l,s)=(1,0)$] is nonreciprocal (reciprocal) for zigzag (straight) shape. However, for $l=0$ and $l=3$, the dispersion relations are symmetric in both zigzag and straight lattice (see Appendix B for details).

\begin{figure*}[!htbp]
\begin{centering}
\includegraphics[width=0.98\textwidth]{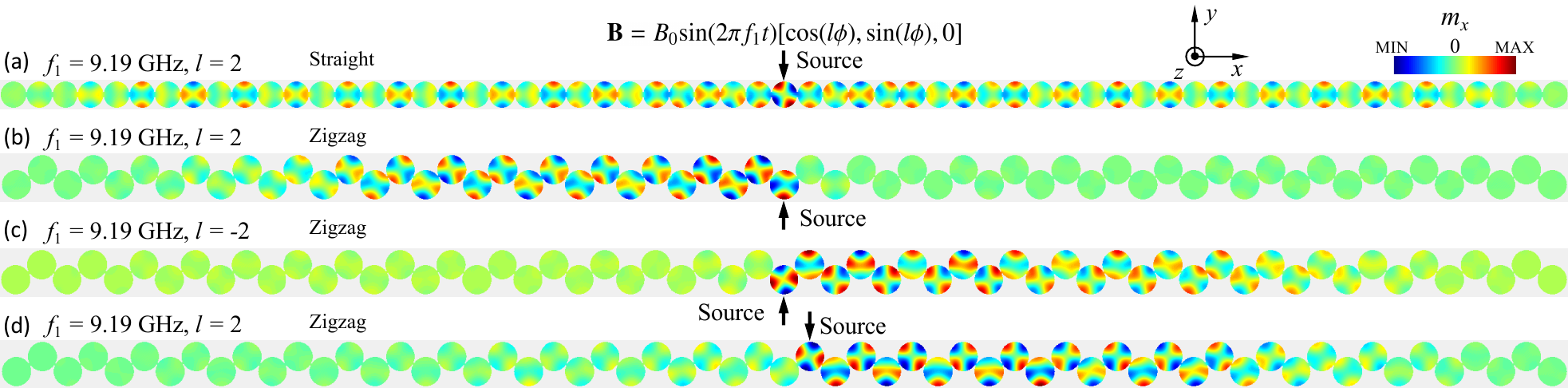}
\par\end{centering}
\caption{The spatial distribution of TM intensity in one-dimensional (a) straight and (b)-(d) zigzag nanodisk arrays. Here $l=2$ in (a), (b), and (d), $l=-2$ in (c). Black arrows denote the disk where the driving field $\mathbf{B}$ is applied. The simulation time $t=10$ ns.}
\label{Figure5}
\end{figure*}

To further visualize the nonreciprocal propagation of TMs, we choose one representative frequency: $f_{1}=9.19$ GHz for $(l,s)=(2,0)$, as marked by red lines in Fig. \ref{Figure3}. We then simulate the dynamics of TMs by the excitation field
\begin{equation}\label{Eq8}
\mathbf{B}(t)=B_{0}\text{sin}(2\pi f_{1}t)[\text{cos}(l\phi),\text{sin}(l\phi),0],
\end{equation}  
with $B_{0}=1$ mT applied at the center disk, indicated by the black arrows in Fig. \ref{Figure5}. Figure \ref{Figure5}(b) shows the propagation of TMs in the zigzag structure, from which one can clearly observe the unidirectional propagation of TMs. For comparison, we also plot the propagation images of TMs in the straight lattice, as shown in Fig. \ref{Figure5}(a), which shows a symmetric spread. Interestingly, we find that for the zigzag structure, the propagation direction of TMs can be reversed by changing the sign of $l$ [see Fig. \ref{Figure5}(c)] or the position of excitation field [see Fig. \ref{Figure5}(d)].

The physical mechanism of the symmetric and asymmetric TM dispersion relations can be explained as a geometric effect. For $l=0$, because the phase structure of TM is symmetric along any radius direction [see Fig. \ref{Figure2}], the propagation of TMs are thus symmetric for both straight and zigzag lattice, as shown in Fig. \ref{Figure6}. It is worth noting that this conclusion is always hold for any value of $\theta$ (here the condition $\pi/3<\theta\leq\pi$ should be satisfied to guarantee that there is no overlap between neighboring nanodisks. If $\theta=\pi/3$, each disk is tangent to the four surrounding disks and the system
 is no longer a simple one-dimensional structure). For $l\neq 0$,  we define $\beta=\pi/l$ to represent the angle between the nearest neighbor azimuthal nodes of TM. At first, we must stress the fact that the TM can spread to adjacent disk only when the contact point is not at the node of TM. Considering that the TM is excited in a nanodisk, when the left contact point is (not) located at the node of TMs, the right contact point is also (not) located at the node, if $\theta$ is the integer multiple of $\beta$. In this case, the dispersion relation is symmetric. However, if $\theta$ is not the integer multiple of $\beta$, the two contact points can not at the node simultaneously, the dispersion relation thus asymmetric. These conclusions can be used to explain our results. On the one hand, for the straight lattice, i.e., $\theta=\pi$, no matter what value $l$ takes, $\theta$ is always an integer multiple of $\beta$. Therefore, the dispersion relations for all $l$ are reciprocal [see Fig. \ref{Figure2}]. On the other hand, for the zigzag structure considered in our paper, i.e., $\theta=2\pi/3$, the situation is different. When $l=1$, $\beta=\pi$, the $\theta=2\pi/3$ is not the integer multiple of $\beta$. Naturally, the dispersion relation is asymmetrical for $(l,s)=(1,0)$ [see Fig. \ref{Figure4}]. We can do the similar analysis for $l=2$, in this case, $\beta=\pi/2$, again, the $\theta$ is not the integer multiple of $\beta$, the dispersion relation is thus asymmetri for $(l,s)=(2,0)$ [see Fig. \ref{Figure3}]. However, when $l=3$, $\beta=\pi/3$, we have $\theta=2\beta$, therefore, the band structure is reciprocal for $(l,s)=(3,0)$ [see Figs. \ref{Figure6}(i) and \ref{Figure6}(l)]. Here the spectra show a little nonreciprocity which originates from the fact that the software MUMAX3 is based on the finite difference method, and the position of contact is thus not a strict point. At last, it is worth noting that the propagation direction of the unidirectional TMs depends on both the sign of $l$ and the position of excitation field [we use $P=1$ ($P=-1$) denotes the excitation field located at the lower (upper) disks]. Concretely, when sgn$(l)$sgn$(P)=1$ (or -1), the TMs propagate leftward (or rightward).

Based on the above analysis, we can easily infer that when one contact point is located at the node, if the other contact point is located at the peak (or trough), the nonreciprocity of TMs reach the maximum. In this case, $\theta=(2n+1)\pi/2l$ with $n=0,1,2,3\cdot\cdot\cdot$ (note the condition $\pi/3<\theta\leq\pi$ should be satisfied simultaneously). We therefore can realize the unidirectional propagation of TMs for any nonzero $l$ by tuning $\theta$.

\section{Discussion and conclusion}\label{section5}

The researches about the twisted magnonics are still in the very initial stage, and a lot of questions and new physicsl phenomena need to be answered and discovered. For example, by constructing the Su-Schrieffer-Heeger \cite{SuPRL1979} and Haldane \cite{HaldanePRL1988} models based on magnetic nanodisks, we can realize the topological edge states of TMs, which may have great potential for designing topologically protected high-capacity communication devices. Besides, the interaction between TMs and various spin texture (for example skyrmion, vortex, and domain wall etc.) also deserves careful investigation, which may leads to peculiar physicsl phenomena, for example, magnetic frequency comb \cite{WangPRL2022}.

To conclude, we have studied the collective excitations of TMs in one-dimensional magnetic nanodisk arrays. For a straight lattice, by performing micromagnetic simulations, we identified multiple symmetric bands which characterize different collective modes of TMs. A theoretical model was proposed to explain the band structure and the results agree well with simulations. For the zigzag structure, we found that the TM dispersion relations for $l=1$ and $l=2$ show obvious nonreciprocity, which do not happen for $l=0$ and $l=3$.    
The propagation characteristics (reciprocal or nonreciprocal) of these bands result from a geometric effect: when $\theta$ is (not) the integer multiple of $\beta$ $(=\pi/l$), the dispersion relation is symmetric (asymmetric). Utilizing this principle, we can achieve unidirectional propagation of TMs with any nonzero $l$. Our work provides a simple and effective method to manipulate the propagation TMs, which should be helpful for designing useful TM devices.

\section*{ACKNOWLEDGMENTS}
We thank Z. Wang and H. Y. Yuan for helpful discussions. This work was supported by the National Key Research and Development Program under Contract No. 2022YFA1402802 and the National Natural Science Foundation of China (NSFC) (Grants No. 12074057, No. 11604041, and No. 11704060). Z.-X.L. acknowledges financial support from the NSFC (Grant No. 11904048) and the Natural Science Foundation of Hunan Province of China (Grant No. 2023JJ40694). X.S.W. was supported by the NSFC (Grants No. 12174093 and No. 11804045) and the Fundamental Research Funds for the Central Universities. X. L. acknowledges the support from the Talent Introduction Program of Chengdu Normal University under Grant No.YJRC2021-14.

\section*{APPENDIX A: THE MAGNETOSTATIC ENERGY OF THE SYSTEM}

\begin{figure*}[!htbp]
\begin{centering}
\includegraphics[width=0.85\textwidth]{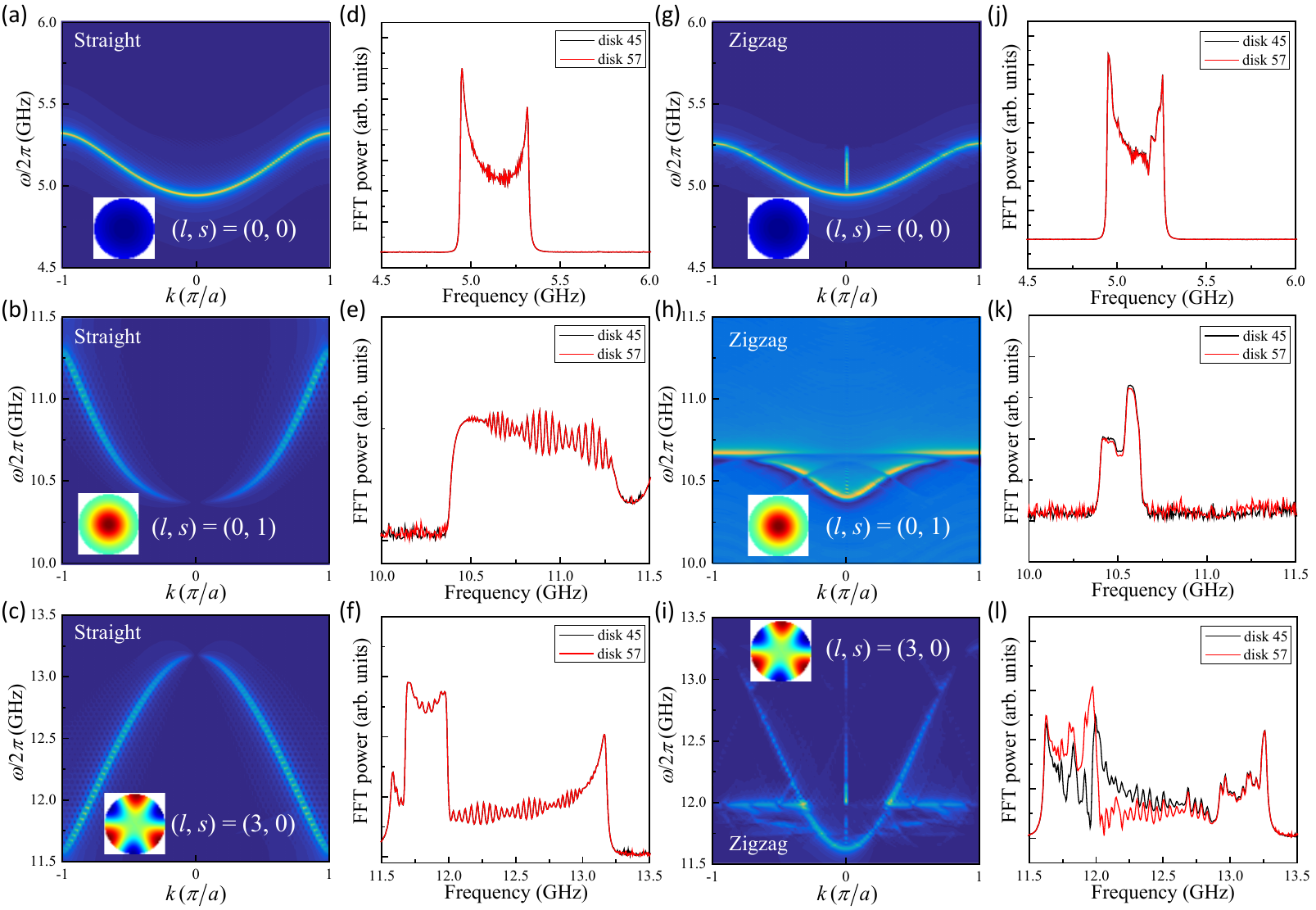}
\par\end{centering}
\caption{The band structures for (a)-(c) straight and (g)-(i) zigzag lattice with different $l$ or $s$ as marked in the images. The corresponding spectra of the magnetization oscillation at disks 45 and 57 for (d)-(f) straight and (j)-(l) zigzag lattice.}
\label{Figure6}
\end{figure*} 

In magnetic nanodisk array, the magnetostatic energy between two arbitrary magnetic moments ($\mathbf{M}_{1}$ and $\mathbf{M}_{2}$) can be expressed as
\begin{equation}\label{Eq9}
\varepsilon(\mathbf{r})=\frac{\mu_{0}}{4\pi}\bigg[\frac{\mathbf{M}_{1}\cdot\mathbf{M}_{2}}{r^{3}}-3\frac{(\mathbf{M}_{1}\cdot\mathbf{r})(\mathbf{M}_{2}\cdot\mathbf{r})}{r^{5}}\bigg],
\end{equation}  
where $\mathbf{r}$ is the position vector from $\mathbf{M}_{1}$ to $\mathbf{M}_{2}$. In our model, all magnetic moments are perpendicularly magnetized along $z$-axis, we thus can define $\mathbf{r}=(r\text{cos}\gamma,r\text{sin}\gamma)$ with $r$ the magnitude of position vector $\mathbf{r}$ and $\gamma$ being the angle between $\mathbf{r}$ and $x$-axis. When TM is excited, $\mathbf{M}_{1}=M_{s}(m_{1}^{x},m_{1}^{y},1)$ and $\mathbf{M}_{2}=M_{s}(m_{2}^{x},m_{2}^{y},1)$, Eq. \eqref{Eq9} can be simplified as
\begin{equation}\label{Eq10}
\varepsilon(\mathbf{r})=\frac{\mu_{0}M_{s}^{2}}{4\pi r^{3}}\Big(1+b\Big),
\end{equation}   
with the parameter $b=m_{1}^{x}m_{2}^{x}(1-3\text{cos}^{2}\gamma)+m_{1}^{y}m_{2}^{y}(1-3\text{sin}^{2}\gamma)-3\text{sin}\gamma\text{cos}\gamma(m_{1}^{x}m_{2}^{y}+m_{1}^{y}m_{2}^{x})$. Under the linear approximation, we have $b=0$, and $\varepsilon(\mathbf{r})=\mu_{0}M_{s}^{2}/4\pi r^{3}$, which means that magnetostatic energy between two arbitrary magnetic moments is a constant. As a result, the whole magnetostatic energy of the system keeps invariant when TMs are excited.

\section*{APPENDIX B: THE BAND STRUCTURES FOR $l=0$ AND $l=3$}

Figure \ref{Figure6} plots the dispersion relation and disk spectra for $l=0$ and $l=3$ with the help of micromagnetic simulations. One can clearly see that the propagations of TMs are absolutely symmetric [see Figs. \ref{Figure6}(a)-\ref{Figure6}(f)] in a straight lattice. For the zigzag lattice, the bands and spectra show symmetric characteristics for $l=0$ [see Figs. \ref{Figure6}(g), \ref{Figure6}(h), \ref{Figure6}(j), and \ref{Figure6}(k)]. There exists a little nonreciprocity for $l=3$ [see Figs. \ref{Figure6}(i) and \ref{Figure6}(l)], which comes from the calculation errors because of the finite difference method (also see related discussions in the main text). We thus conclude that the propagations of TMs are reciprocal for $l=0$ and $l=3$ in both straight and zigzag structure.

\end{document}